\documentclass[%
 reprint,
 superscriptaddress,
 amsmath,amssymb,
 aps,
 pra,
]{revtex4-2}
\usepackage{graphicx}

\usepackage[version = 4]{mhchem} 
\usepackage[dvipsnames]{xcolor} 
\usepackage{bm}
\usepackage{siunitx}
\usepackage{xr}
\newcommand{\eref}[1]{Eq.~\ref{#1}}
\newcommand{\fref}[1]{Fig.~\ref{#1}}

\graphicspath{ {./figures/} }

\begin{document}

\title{Tunable turning point of the frequency--temperature response in suspended bilayer \ce{SiN_x}/\ce{Al} nanofilms}

\author{Yanli Shi}
  \affiliation{Shandong Institute of Advanced Technology, Jinan, Shandong 251000, China}
\author{Hui Wu}
  \affiliation{Shandong Institute of Advanced Technology, Jinan, Shandong 251000, China}
\author{Guangyu Li}
  \affiliation{Institute of Thermal Science and Technology, Shandong University, Jinan, Shandong 250061, China}
\author{Zhuo Miao}
  \affiliation{Institute of Advanced Technology, Shandong University, Jinan, Shandong 250061, China}
\author{Luyao Zhu}
  \affiliation{Shandong Institute of Advanced Technology, Jinan, Shandong 251000, China}
\author{Chen Zheng}
  \affiliation{Shandong Institute of Advanced Technology, Jinan, Shandong 251000, China}
\author{Weidong Zheng}
  \affiliation{Shandong Institute of Advanced Technology, Jinan, Shandong 251000, China}
\author{Qun Cao}
  \thanks{corresponding authors: Qun Cao (qun.cao@iat.cn) and Zhe He (zhe.he@iat.cn)}
  \affiliation{Shandong Institute of Advanced Technology, Jinan, Shandong 251000, China}
\author{Zhe He}
  \thanks{corresponding authors: Qun Cao (qun.cao@iat.cn) and Zhe He (zhe.he@iat.cn)}
  \affiliation{Shandong Institute of Advanced Technology, Jinan, Shandong 251000, China}
\author{Zheng Cui}
  \affiliation{Shandong Institute of Advanced Technology, Jinan, Shandong 251000, China}


\begin{abstract}
  Non-invasive thermometry is highly desirable in cryogenic experiments to avoid the self-heating inherent in conventional sensors.
  In previous work, temperature was probed via the vibrational eigenfrequency of a suspended single-layer \ce{SiN_x} nanofilm.
  However, the turning point of the frequency--temperature curve limited the usable measurement range.
  Moreover, the slow thermal response confined measurements to a stepwise protocol with only discrete temperature points.
  Here, we show that these limitations can be overcome with a suspended bilayer \ce{SiN_x}/\ce{Al} nanofilm, whose temperature-dependent vibrational eigenfrequency provides a monotonic, continuous, and repeatable temperature readout.
  The frequency--temperature dependence is governed by the mismatch between the effective coefficients of thermal expansion (CTE) of the bilayer and the substrate, which makes the turning point tunable through the thickness ratio of the two layers.
  Over the \qtyrange{90}{300}{K} range, we characterize three bilayer nanofilms with different \ce{Al} layer thicknesses and show how the turning point shifts with the thickness of the \ce{Al} layer.
 For the sample with \qty{72}{nm} \ce{Al}, the turning point is pushed beyond \qty{300}{K}, yielding a monotonic negative temperature coefficient over the entire \qtyrange{90}{300}{K} range.
  Combining a thin \ce{Al} layer on the \ce{SiN_x} film with improved thermal anchoring of the chip shortens the thermal response to below \qty{1}{s}, enabling continuous frequency tracking at a ramp rate of \qty{3}{K/min}.
  We also demonstrate that the nanofilm with a \qty{72}{nm} \ce{Al} layer achieves a repeatability better than \qty{150}{mK} above \qty{175}{K}, and the frequency resolution of \qty{1}{Hz} corresponds to a temperature resolution of $\sim\!\qty{5}{mK}$.
\end{abstract}


\maketitle

\section{Introduction}

Experiments at cryogenic temperatures require precise temperature measurement and control.
Accurate, local measurement of temperature is essential in both fundamental physics and engineering applications, ranging from superconducting quantum computation~\cite{Anferov24Schuster} to astronomical photon detection~\cite{Wang25Vega,Nagler21Wollack}.
In superconducting circuits and single-photon detectors, for example, qubit coherence times~\cite{Burnett19Bylander,Lisenfeld10Ustinov} and dark-count rates~\cite{Hofbauer18Zimmermann,Murphy15Bezryadin} are highly sensitive to thermal fluctuations, while the frequent write and readout operations themselves generate heat~\cite{Lecocq21Teufel,Yamamoto24Tokura,Arnold25Fink,Chowdhury26Oliver}.
Temperature must therefore be tracked locally and in real time, with minimal influence on the measured system.

Most cryogenic thermometers are electrical, with resistive elements~\cite{Childs00Long} and semiconductor diodes~\cite{Courts02Swinehart} are the most widely deployed~\cite{Yeager01Courts} and are favored for their low cost and compact size.
However, these electrical thermometers are invasive by construction.
The sensing current dissipates Joule heat, with the self-heating most severe at the lowest temperatures, where heat capacities are small.
This motivates non-invasive thermometry that adds neither wiring nor dissipation to the sample.

In our previous work, temperature was probed via the vibrational eigenfrequency of a suspended single-layer \ce{SiN_x} nanofilm~\cite{Li26Cui}.
This implementation had two limits.
First, the frequency--temperature curve exhibited a turning point, set by the CTE mismatch between the nanofilm and the substrate.
The temperature sensitivity vanishes at the turning point, and beyond it the temperature coefficient changes sign, so that two different temperatures map to the same frequency.
Second, the thermal response was slow.
Each temperature point required minutes of acquisition, forcing a stepwise protocol where only discrete temperatures were measured and transient thermal processes could not be tracked.

In this work, we study the temperature-dependent vibrational eigenfrequency of a suspended heterogeneous (bilayer) nanofilm and show that the bilayer scheme overcomes these limits.
First, the effective CTE of the heterogeneous film can be engineered through the thickness ratio of the two layers, so that the turning point can be tailored to a target temperature by selecting $h_2$.
For the \qty{72}{nm} aluminum layer, the turning point is pushed beyond \qty{300}{K}, yielding a monotonic negative temperature coefficient over the entire \qtyrange{90}{300}{K} range.
Second, the nanofilm chip is glued to a copper ring that is clamped to the cold finger, improving the thermal anchoring.
The high-conductivity metal layer on the \ce{SiN_x} film further improves heat spreading.
The thermal response is shortened to below the \qty{1}{s} PSD-acquisition time.
The eigenfrequency can thus be tracked continuously during a \qty{3}{K/min} temperature ramp, replacing the previous stepwise protocol.
Moreover, the nanofilm is driven by thermal noise in the high-vacuum, cryogenic environment, so that neither external actuation nor a phase-locked loop (both of which would introduce additional noise) is required.
The temperature dependence of the bilayer film eigenfrequency derived from the CTE-mismatch model agrees well with the frequency--temperature curves measured on three samples with different thickness ratios.
The effective CTEs extracted from the measurements follow the theoretical predictions, and the resulting shift of the turning point demonstrates the potential of the bilayer scheme for monotonic, non-invasive thermometry.

\section{Theory}

We extend the analysis of our previous single-layer work~\cite{Li26Cui} to the bilayer configuration.
As illustrated in \fref{fig:exp_setup}~(a), the bilayer nanofilm consists of two firmly attached layers with the same lateral size $l$ and respective thicknesses $h_i$, where the subscript $i=1,2$ denotes the first and second layer.
Within the thin-film approximation $l^2\gg h^2E/\sigma$ (with $\sigma$ and $E$ the stress and Young's modulus of the nanofilm), the vibration of the square bilayer nanofilm is governed by
\begin{align}
    \tau \nabla^2 \bm{r}
    &=
    \mu \frac{\partial^2}{\partial t^2} \bm{r}
    ,
\end{align}
where $\bm{r}$ is the displacement of the bilayer nanofilm, and $\tau=\sum h_i\sigma_i$ and $\mu=\sum h_i\rho_i$ are the tension and the area density, with $\rho_i$ the mass density of layer $i$.
With fixed boundary conditions, the vibrational eigenfrequency is
\begin{align}
    f_{(m,n)}
    &=
    \frac{\sqrt{m^2+n^2}}{2l}
    \sqrt{\frac{\tau}{\mu}}
    ,
\end{align}
where $m,n=1,2,3,\cdots$ are the mode indices of the vibration.
The temperature dependence of the squared fundamental mode eigenfrequency is governed by the CTE mismatch between the effective CTE of the bilayer and that of the substrate,
\begin{align}
    \frac{\partial f_{(1,1)}^2}{\partial T}
    &=
    -
    \frac{M_1}{2l^2\rho'}
    \left(\alpha'-\alpha_\text{s}\right)
    \label{eq:freq_temp_dependency}
    ,
\end{align}
where the thermal expansion of the lateral size and of the areal density has been neglected relative to that of the tension, and
\begin{align}
    \rho'
    &=
    M_1\frac{h_1\rho_1+h_2\rho_2}{h_1M_1+h_2M_2}
    ,
    \\
    \alpha'
    &=
    \frac{h_1M_1\alpha_1+h_2M_2\alpha_2}{h_1M_1+h_2M_2}
    \label{eq:effective_CTE}
    ,
\end{align}
with $\alpha'$ the effective CTE of the bilayer system, $M_i=E_i/(1-\nu_i)$ the biaxial modulus, $\nu_i$ the Poisson's ratio, $\alpha_i$ the CTE of layer $i$, $\alpha_\text{s}$ the CTE of the substrate, and $\rho'$ an auxiliary quantity defined so that $M_1/\rho'=(\sum_i h_iM_i)/(\sum_i h_i\rho_i)$.

The frequency--temperature curve exhibits a turning point at $T_\text{turn}$, where the effective CTE of the bilayer equals that of the substrate ($\alpha'=\alpha_\text{s}$).
At $T_\text{turn}$ the temperature sensitivity $\partial f_{(1,1)}/\partial T$ vanishes, making the turning point a blind spot that limits the usable measurement range.
For a \ce{SiN_x}--metal bilayer on a \ce{Si} substrate, increasing the metal layer thickness $h_2$ raises $\alpha'$ over the entire temperature range, shifting the turning point to higher temperatures.
The turning point can thus be tuned through $h_2$ and pushed outside the operating range of interest.

\section{Experimental Setup}

\begin{figure*}[t!]
  \centering
  \includegraphics[width=0.75\linewidth]{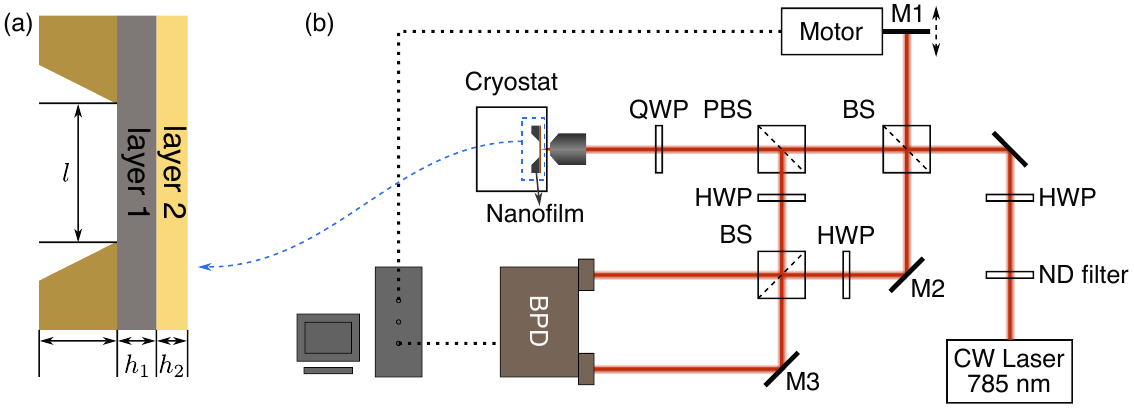}
  \caption{\label{fig:exp_setup}
    (a)
    Schematic of the bilayer nanofilm.
    A first layer of thickness $h_1$ is deposited on the substrate, and a second layer of thickness $h_2$ on top of the first.
    (b)
    Schematic of the Mach-Zehnder interferometer used to track the eigenfrequency of the suspended nanofilm as the temperature changes.
    The laser beam is split into a signal arm, which traverses the cryostat holding the nanofilm, and a reference arm, which is steered by mirrors M1 and M2.
    A motorized mirror M1, driven by a PID loop, locks the interferometer at maximum sensitivity.
    The interference signal is recorded by a balanced photodetector~(BPD), and the fundamental eigenfrequency is obtained from its power spectral density.
    CW, continuous-wave; ND, neutral density filter; HWP, half-wave plate; QWP, quarter-wave plate; BS, beam splitter; PBS, polarizing beam splitter; M, mirror.
  }
\end{figure*}

\begin{figure}[htb!]
    \centering
    \includegraphics[width=\linewidth]{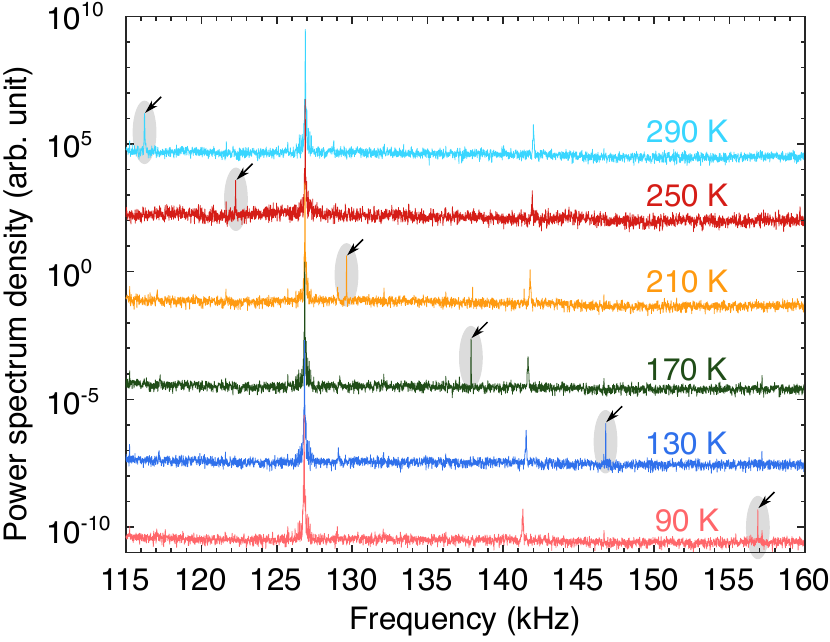}
    \caption{\label{fig:raw_data}
    Power spectral density (PSD) of the raw data from sample C for selected temperatures.
    The $(1,1)$-mode eigenfrequency $f_{(1,1)}$ of sample C (marked by arrows) drops as temperature increases.
    Two stationary peaks, located at about \qtylist{127;142}{kHz}, barely move with temperature and are attributed to vibrations of the rest of the cryostat and to the laser source.
    PSD profiles at different temperatures are vertically offset for clarity.
    }
\end{figure}

The thermal vibration of the suspended bilayer nanofilm is probed by a Mach-Zehnder interferometer with balanced detection, as illustrated in \fref{fig:exp_setup}.
The bilayer nanofilm is made by thermal evaporation of an aluminum~(\ce{Al}) layer on a suspended \ce{SiN_x} nanofilm.
The chip is glued onto an oxygen-free copper (TU0) ring, and the ring is clamped onto the temperature-controlled cold finger inside a liquid-nitrogen cryostat, with indium~(In) foil inserted between the ring and the cold finger to improve the thermal connection.
The cryostat covers \qtyrange{80}{300}{K} with a temperature stability better than \qty{1}{mK}.

A \qty{785}{nm} continuous-wave laser beam is divided into a signal arm and a reference arm by a beam splitter~(BS).
In the signal arm, a polarizing beam splitter~(PBS) and a quarter-wave plate~(QWP) direct the beam onto the nanofilm inside the cryostat and, after the double pass through the QWP, route the reflected beam toward the detector.
In the reference arm, the beam is steered by mirrors M1 and M2, with half-wave plates~(HWPs) adjusting its polarization.
The two arms are recombined at a second BS, and the resulting interference is recorded by a balanced photodetector~(BPD).
Active stabilization is provided by the motor-controlled mirror~(M1), whose position is adjusted by a software PID loop that locks the interferometer at its point of maximum sensitivity.
From the detector signal, the power spectral density~(PSD) is computed and the fundamental eigenfrequency $f_{(1,1)}$ is tracked with a \qty{1}{s} acquisition per spectrum as the temperature varies (\fref{fig:raw_data}).

\section{Results}\label{sec:results}

\begin{figure}[tb!]
    \centering
    \includegraphics[width=\linewidth]{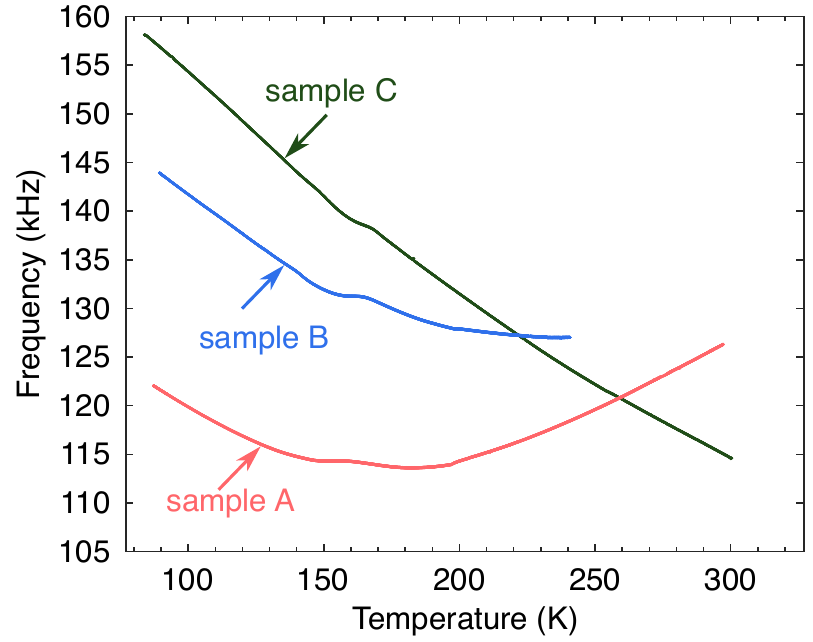}
    \caption{\label{fig:turning_point}
    Eigenfrequency of samples A ($h_2=\qty{24.6}{nm}$), B ($h_2=\qty{41.6}{nm}$) and C ($h_2=\qty{72.0}{nm}$) as a function of temperature.
    Increasing the aluminum-layer thickness shifts the turning point of the frequency--temperature curve to higher temperatures, from $\sim\qty{170}{K}$ (sample A) to beyond \qty{300}{K} (sample C).
    The vertical offset among samples arises from differences in residual stress and does not affect the turning-point behavior.
    Each trace spans the temperature range over which the $(1,1)$-mode peak was tracked for that sample.
    }
\end{figure}

The first layer of the bilayer is a commercial non-stoichiometric silicon nitride~(\ce{SiN_x}) nanofilm (Norcada Inc., NX5100D) with a measured thickness $h_1=\qty{205.2}{nm}$, onto which the aluminum layer is deposited by thermal evaporation.
Three samples are investigated in this work, with \ce{Al} thickness $h_2=\qty{24.6}{nm}$ (sample A), \qty{41.6}{nm} (sample B), and \qty{72.0}{nm} (sample C).
The lateral size of the films is measured to be $l=\qty{1.026}{mm}$.
The material parameters used in the model are
$E_\text{\ce{SiN_x}}=\qty{263}{GPa}$~\cite{Chuang04Ghodssi},
$E_\text{Al}=\qty{69}{GPa}$~\cite{Chinmulgund95Barnard},
$\nu_\text{\ce{SiN_x}}=0.23$~\cite{Edwards04Sharpe},
$\nu_\text{Al}=0.33$~\cite{Chinmulgund95Barnard},
$\rho_\text{\ce{SiN_x}}=\qty{3.17}{g.cm^{-3}}$~\cite{Haynes17},
$\rho_\text{Al}=\qty{2.70}{g.cm^{-3}}$~\cite{Arblaster18},
$\alpha_\text{Si}(T)$ from reference~\cite{Middelmann15Schodel},
$\alpha_\text{Al}(T)$ from reference~\cite{Arblaster18}, and
$\alpha_\text{\ce{SiN_x}}(T)$ measured experimentally with a single-layer nanofilm.

Unlike the stepwise protocol used in our previous paper~\cite{Li26Cui}, we measured the frequency--temperature dependence continuously, with the cryostat temperature increasing at \qty{3}{K/min}.
To validate the continuous approach, we performed a thermal response test with stepwise temperature changes, as shown in Fig.~S1. 
At each step, the set point was changed by a PID controller (Dexin TC290) and the temperature was then held for at least \qty{300}{s} to allow thermal equilibration.
Fig.~S1 and Fig.~S2 show that the thermal response time of the nanofilm is shorter than \qty{1}{s}, both for the large temperature step at $t=t_\text{rise}$ and for the small residual changes over $\qty{1550}{s}\lesssim t\lesssim\qty{1800}{s}$. 
The nanofilm therefore thermalizes faster than a single \qty{1}{s} PSD acquisition, so each spectrum reflects the instantaneous temperature of the film.
The resolution of the PSD spectrum is about \qty{1}{Hz}, which, at the local slope $|\partial f_{(1,1)}/\partial T|\approx\qty{200}{Hz/K}$, corresponds to a temperature resolution of about \qty{5}{mK}.

To check the influence of laser-induced heating of the nanofilm, we varied the incident laser power between \qty{1}{\micro\watt} and \qty{5}{\micro\watt} while the temperature was held at \qty{140}{K}.
The eigenfrequency changed by less than \qty{30}{Hz} (Fig.~S3).

Increasing the thickness of the second layer shifts the turning point of the frequency--temperature curve to higher temperatures, as shown in \fref{fig:turning_point}.
For sample A ($h_2=\qty{24.6}{nm}$), the eigenfrequency first decreases and then increases with temperature, with a turning point at $T_\text{turn}\approx\qty{170}{K}$.
For sample B ($h_2=\qty{41.6}{nm}$), the turning point shifts to $\approx\qty{240}{K}$.
For sample C ($h_2=\qty{72.0}{nm}$), it is pushed beyond \qty{300}{K}, so that the negative temperature coefficient spans the entire measured range.

As shown in \fref{fig:turning_point}, a frequency drift appears between \qty{150}{K} and \qty{170}{K}.
The drift occurs in all three samples, regardless of the second-layer thickness.
Replacing the \ce{Al} layer with a different metal such as \ce{Au} does not remove it. 
Whenever the temperature is cycled through this range, the repeatability of the frequency--temperature curve is degraded.
The repeatability is therefore evaluated above this range, where it approaches the intrinsic value set by the bilayer film itself.
We attribute this anomalous drift to the mounting of the sample.

Furthermore, we quantify the short-term repeatability of the sensor by comparing the eigenfrequency traces of two measurement runs on sample C, as shown in Fig.~S4. 
Over the range \qtyrange{175}{260}{K} above the anomalous-drift region discussed above, the absolute difference between the two runs remains below \qty{30}{Hz}, corresponding to a temperature discrepancy below \qty{150}{mK}.

\begin{figure}[tb!]
    \centering
    \includegraphics[width=\linewidth]{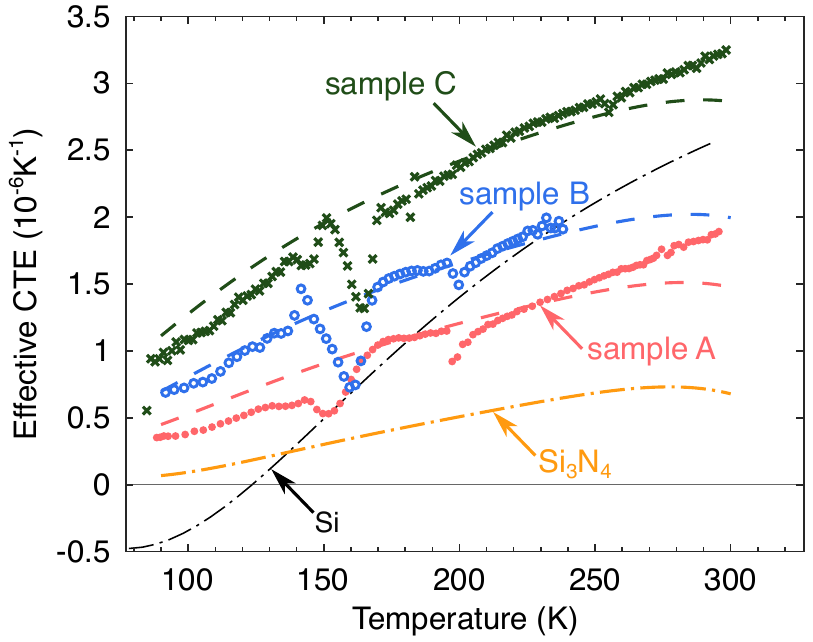}
    \caption{\label{fig:CTE_manipulation}
    Effective CTEs of samples A, B, and C (filled red circles, open blue circles, and dark green crosses, respectively).
    The dot-dashed lines indicate the CTEs of the silicon substrate (black) and the \ce{SiN_x} nanofilm (yellow), and the dashed lines show the theoretical values for each sample.
    Each data point is obtained from a linear fit to 100 raw PSD profiles.
    The \ce{SiN_x} CTE is estimated experimentally with a single-layer nanofilm.
    It is fitted to a fifth-order polynomial.
    The \ce{Al} CTE (\qtyrange{12e-6}{23e-6}{K^{-1}}) lies above the plotted range~\cite{Arblaster18}.
    }
\end{figure}

The turning point is tuned by engineering the effective CTE of the bilayer nanofilm.
As \eref{eq:effective_CTE} shows, the effective CTE can be adjusted through the thickness or the material of the second layer.
Aluminum was selected as the second-layer material because its CTE substantially exceeds that of \ce{SiN_x} throughout the measured range ($\alpha_\text{\ce{Al}}(\qty{100}{K})\approx\qty{12e-6}{K^{-1}}$ and $\alpha_\text{\ce{Al}}(\qty{300}{K})\approx\qty{23e-6}{K^{-1}}$)~\cite{Arblaster18}, raising $\alpha'$ above $\alpha_\text{s}$ so that the eigenfrequency decreases with temperature over the entire range, as follows from \eref{eq:freq_temp_dependency}.
The predicted effective CTEs of all three samples are shown as dashed lines in \fref{fig:CTE_manipulation}.

Inverting \eref{eq:freq_temp_dependency}, the effective CTE of the bilayer system can be extracted from the measured frequency,
\begin{align}
  \alpha'
  &=
  \alpha_\text{s}
  -
  \frac{2l^2\rho'}{M_1}
  \frac{\partial f^2_{(1,1)}}{\partial T}
  \label{eq:effective_CTE_from_freq}
  .
\end{align}
The extracted effective CTEs are also plotted in \fref{fig:CTE_manipulation} as filled circles (sample A), open circles (sample B), and crosses (sample C).
They agree well for all three samples.
In particular, the predicted crossing with the substrate CTE is reproduced, confirming that the bilayer behaves as a single film with a stiffness- and thickness-weighted CTE.
Near \qty{150}{K}, the extracted $\alpha'$ deviates from the theoretical curves, reflecting the drift observed in \fref{fig:turning_point}.

\section{Conclusion}\label{sec:conclusion}

In summary, we have demonstrated a non-invasive cryogenic thermometer based on the temperature-dependent vibrational eigenfrequency of a suspended bilayer \ce{SiN_x}/\ce{Al} nanofilm, whose temperature coefficient is tunable through the thickness ratio of the two layers.
The effective-CTE model agrees well with the measured frequency--temperature curves, and the turning point shifts from $\sim\qty{170}{K}$ for $h_2=\qty{24.6}{nm}$ to beyond \qty{300}{K} for $h_2=\qty{72.0}{nm}$, so that for the thickest sample the blind spot falls outside the measured range.
The improved thermal anchoring yields a sub-second thermal response, enabling continuous tracking during \qty{3}{K/min} temperature ramps, and the readout reaches a frequency resolution of \qty{1}{Hz} ($\sim\!\qty{5}{mK}$) with a short-term repeatability better than \qty{150}{mK} above \qty{175}{K}.
These advances overcome the two limitations of the previous single-layer implementation, namely the fixed turning point and the slow thermal response.
The nanofilms exhibit an anomalous frequency drift near \qty{150}{K}, which we attribute to the mounting of the sample.
Eliminating this drift is a prerequisite for practical thermometry below \qty{170}{K} and is left for future work.

\begin{acknowledgments}
The project is supported by the National Natural Science Foundation of China under grant No. 12404413, Natural Science Foundation of Shandong Province (No. 2024HWYQ-081, ZR2026MS0050), and Taishan Scholars Foundation of Shandong Province (No. tsqn202408307, No. tsqn202306316).
Chen Zheng acknowledges the support from the National Natural Science Foundation of China under grant No. 12605319.
\end{acknowledgments}

\section*{Data Availability}
The data and analysis code of this study are available from the corresponding authors upon reasonable request.

\bibliographystyle{apsrev4-2}
\bibliography{frequency_thermometer_reference}

\end{document}